\documentclass[final,3p,times,twocolumn]{elsarticle} %

\usepackage{hyperref}

\usepackage{graphicx}
\usepackage{amssymb}
\usepackage{lineno}

\biboptions{comma,square}

\journal{Nucl. Instr. Meth. Phys. Res. A}

\begin{document}

\begin{frontmatter}



\title{The C-4 Dark Matter Experiment}


\author[PNNL]{R.M.~Bonicalzi}
\author[UC]{J.I.~Collar\corref{cor}}\ead{collar@uchicago.edu}
\author[Canberra]{J.~Colaresi}
\author[PNNL]{J.E.~Fast}
\author[UC]{N.E.~Fields}
\author[PNNL]{E.S.~Fuller}
\author[UC]{M.~Hai}
\author[PNNL]{T.W.~Hossbach}
\author[PNNL]{M.S.~Kos}
\author[PNNL]{J.L.~Orrell\corref{cor}}\ead{john.orrell@pnnl.gov}
\author[PNNL]{C.T.~Overman}
\author[PNNL]{D.J.~Reid}
\author[PNNL]{B.A.~VanDevender}
\author[PNNL]{C.~Wiseman\corref{USC}}
\author[Canberra]{K.M.~Yocum}

\cortext[cor]{Corresponding authors.}
\cortext[USC]{Current address: Department of Physics and Astronomy, University of South Carolina, Columbia, South Carolina, 29208 USA}

\address[PNNL]{Pacific Northwest Laboratory, Richland, WA 99352, USA}
\address[UC]{Kavli Institute for Cosmological Physics and Enrico Fermi Institute, University of Chicago, Chicago, IL 60637, USA}
\address[Canberra]{CANBERRA Industries, Meriden, CT 06450, USA}

\begin{abstract}
We describe the experimental design of C-4, an expansion of the CoGeNT dark matter search to four identical detectors each approximately three times the mass of the p-type point contact germanium diode presently taking data at the Soudan Underground Laboratory. Expected reductions of radioactive backgrounds and energy threshold are discussed, including an estimate of the additional sensitivity to low-mass dark matter candidates to be obtained with this search.
\end{abstract}

\begin{keyword}


dark matter experiment \sep direct detection \sep low-mass dark matter \sep CoGeNT \sep low threshold germanium ionization spectrometer

\end{keyword}

\end{frontmatter}


%\linenumbers


\section{Introduction}
\label{Introduction}

The C-4 dark matter experiment is intended to test recent results from the CoGeNT experiment \cite{longpaper}. CoGeNT (Coherent Germanium Neutrino Technology), located at the Soudan Underground Laboratory (SUL), reported an excess of events above the estimated background, for events having electron scattering equivalent energies of less than 3~keVee (keV electron equivalent (keVee), i.e., ionization energy) \cite{cogentPRL106}. Following 15-months of data collection, a low-significance (2.8$\sigma$) modulation in the low-energy rate was found to have a roughly annual period and a phase peaking in summer \cite{cogentPRL107}, not unlike the expectation for an annual modulation induced by dark matter interactions \cite{Drukier86}. These results from the CoGeNT experiment may be interpreted as either evidence for a low-mass (5-10~GeV/c$^{2}$) dark matter particle, or an unidentified systematic effect or background. These results call for a follow-up experiment with improved sensitivity able to clarify the situation. This improved sensitivity stems from increasing the target mass, lowering the background, and potentially reducing the energy threshold, as described in this article.

\begin{figure}[ht!]
\begin{center}
\includegraphics[width=\columnwidth]{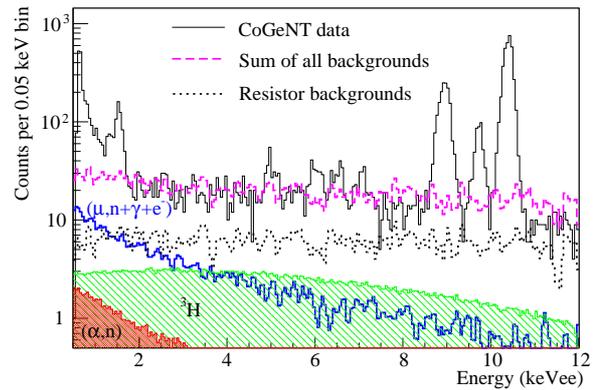}
\caption{\label{fig:cogentbkgrnd} Background estimation for the CoGeNT experiment \cite{longpaper}. This estimate is used as a basis for addressing background reduction in the C-4 experimental design.}
\end{center}
\end{figure}

The CoGeNT experiment employed a single 440-gram germanium crystal of the p-type point contact (PPC) design \cite{JCAP}. The C-4 dark matter experiment will employ four similarly designed germanium spectrometers in individual cryostats, co-located in a single underground low-background shield. These detectors will have at minimum the same performance characteristics of the CoGeNT detector \cite{cogentPRL106}, namely $\sim$0.5~keVee energy threshold and $\sim$235~eV full-width at half-maximum (FWHM) energy resolution at 5.9~keVee. However, C-4 will employ larger germanium crystals, up to 1.3~kg per crystal, resulting in a factor of $\sim$10 greater active mass than CoGeNT directly providing increase sensitivity from total target exposure. Figure~\ref{fig:cogentbkgrnd} is reproduced from an analysis of the CoGeNT background \cite{longpaper} and is the basis for understanding how to address background reduction through improved detector and cryostat design for the C-4 experiment. Specifically, Figure~\ref{fig:cogentbkgrnd} shows an improved cosmic ray veto, better neutron shielding, and reduction of ubiquitous radioisotopes in the electronics would improve the background of the experiment. Additionally, work is underway to further reduce the energy threshold in the next generation of $\sim$1~kg detectors (Section \ref{sec:noise}). Operation of multiple detectors will test the reproducibility of the unexpected spectral features observed by CoGeNT below 3~keVee. A search for a dark matter-induced annual modulation of the event rate will benefit from the consistency check provided by inter-comparison of the four C-4 detectors, and the improved statistical evidence brought by the larger target mass (Section \ref{sec:science}). In summary, the C-4 dark matter experiment seeks to test the CoGeNT results through greater statistical sensitivity, improved shielding, lower backgrounds, potentially a reduced energy threshold, and detector-to-detector consistency checks. This article presents the design of the C-4 dark matter experiment, followed by background estimation and potential reduction of the detector's energy thresholds. Finally, the resulting expected sensitivity of the experiment is presented.

\section{Experimental Design}
\label{ExperimentalDesign}

\begin{figure}[ht!]
\begin{center}
\includegraphics[width=\columnwidth]{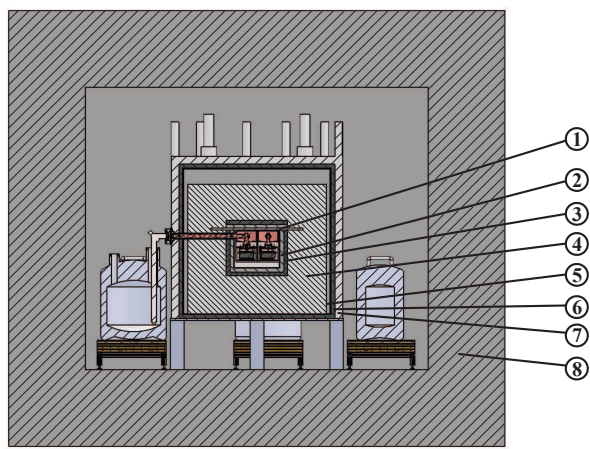}
\caption{\label{fig:ExpDesign} Cross-section of the C-4 shield design. The eight numbered shield layers (thicknesses following each in parenthesis) include \textbf{(1)} oxygen-free high conductivity (OFHC) copper cage for internal dimensional control (3.18~mm) and OFHC copper sheet to support lead above the cavity (12.7~mm), \textbf{(2)} ultra-low background lead (25.4~mm), \textbf{(3)} low background lead (25.4~mm), \textbf{(4)} stock lead (254~mm), \textbf{(5)} aluminum radon exclusion box (3.18~mm), \textbf{(6)} 30\% borated polyethylene (25.4~mm), \textbf{(7)} plastic scintillator anti-cosmic veto (50.8~mm), and \textbf{(8)} water / high density polyethylene (HDPE) neutron moderator (508~mm~/~610~mm).}
\end{center}
\end{figure}

The C-4 shield design is conceptually similar to that for the CoGeNT detector \cite{longpaper}. The present plan is to locate the C-4 dark matter experiment in the proton-decay hall of the Soudan Underground Laboratory (2800~m.w.e. overburden), the same location as the CoGeNT and CDMS-II experiments. Figure \ref{fig:ExpDesign} shows a cut-away cross-section of the C-4 shield design including descriptions and thicknesses of the various layers. In contrast to the CoGeNT shield, the C-4 shield increases the thickness of the lead shield by 50.8~mm and adds a thick plastic scintillator anti-cosmic veto. Furthermore, the CoGeNT experiment's neutron moderating outer shield was composed of a 183~mm think HDPE layer, while the C-4 shield design more than doubles this thickness via a combination of HDPE (above and below) and water-filled HPDE containers (vertical walls).%, similar to those used in the COUPP-4kg experiment at SNOLAB.

\subsection{Data Acquisition System}

The data acquisition system used on the CoGeNT experiment is a hybrid design that combines digital and analog electronics \cite{longpaper}. In CoGeNT, the trigger is generated from a level discriminator applied to the shaped output of a spectroscopy amplifier, and all channels are recorded using waveform digitizers, for subsequent post-processing analysis. The data acquisition system planned for the C-4 experiment relies entirely on a fast digitizer able to record raw preamplifier traces and to trigger on individual channels, corresponding to each of the four detectors. The triggering conditions are specified via field-programmable gate array (FPGA) logic, able to filter out noise transients dissimilar to a radiation-induced pulse. The need for this ``intelligent'' triggering arises from the realization that a majority of noise triggers below threshold in the CoGeNT detector originate from environmental transients in electromagnetic/radio frequency (EM/RF) noise, not on noise intrinsic to the detector. These pulses can be easily rejected offline, but they limit the lowest energy detectable by the CoGeNT detector, if a reasonable triggering rate is to be preserved. We expect a significant reduction in the energy threshold in C-4 detectors, presently at $\sim$0.45~keVee for CoGeNT, through the adoption of FPGA-based triggering. A second reason for the data acquisition upgrade is the isolation of any unknown source of instability in the present system that might lead to seasonal modulations in detection efficiency \cite{longpaper}. 

The heart of the C-4 data acquisition system is the four-channel National Instruments NI~5734 16-bit, 120~MS/s, high-speed digitizer module, used in combination with a NI PXIe-7966R FlexRIO FPGA module. Additional slow data acquisition modules provide the capability for analog and digital signal collection of other monitoring equipment (e.g., temperature, humidity, radon sampling, seismic activity, etc.) and control of automatic liquid nitrogen (LN) dewar refills \cite{longpaper}.  Bias to the detectors is provided by a CAEN N1471 four-channel programmable high-voltage supply. Tests of trigger efficiency near threshold and its dependence on pulse rise time will be performed with an Agilent 33521A arbitrary waveform generator.

\subsection{Detector Characterization}

A full characterization of the active volume of each of the C-4 germanium detectors is of crucial importance. Each PPC detector features an inner active bulk volume where complete charge collection takes place. The exterior surface of each germanium detector is an inactive ``dead'' layer. This layer originates in the lithium diffusion necessary to create an outer contact, and is typically of order $\sim$1~mm thickness. Between the dead-layer and the active bulk is a transition region of similar thickness where charge collection (and measured energy) is only partial. Pulses corresponding to charge depositions in this transition region have a characteristic slow rise-time \cite{longpaper} and degraded energy estimation. The goal of the planned characterization of each C-4 detector is to quantify the effect of the transition region as well as establishing the total active (bulk) volume of the detectors.

The characterization will include irradiation with suitable collimated and uncollimated low-energy gamma sources able to help determine the inactive dead-layer thickness and its uniformity \cite{longpaper}. Detector scanning and characterization will take place during the ``cool-down'' period following installation underground, a period of time over which short-lived ($\sim$few weeks half-life) cosmogenically produced isotopes decay away. Short-lived cosmogenic isotopes leading to low-energy depositions (e.g., $^{71}$Ge) can be used to study the impact of the transition region on the accumulation of events near threshold \cite{longpaper}.

\section{Background Estimates}
\label{sec:backgrounds}

\begin{figure}[ht!]
\begin{center}
\includegraphics[width=\columnwidth]{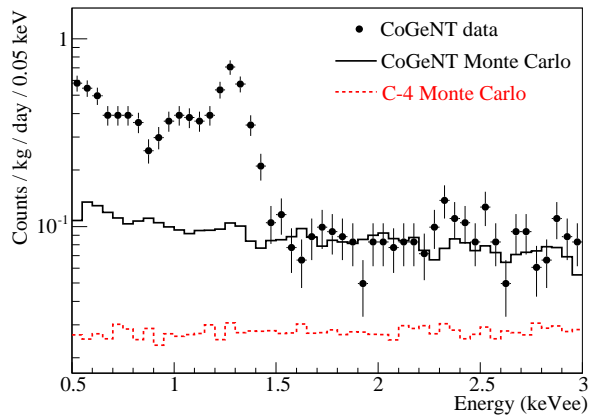}
\caption{\label{fig:comparison} Comparison of C-4's predicted continuum backgrounds to CoGeNT in the low-energy region. The CoGeNT data (points) includes a contribution from L-shell electron capture (peaks around 1.0--1.3~keV) stemming from cosmogenic activation in germanium. These peaks are not shown for the Monte Carlo C-4 background prediction (red dashed line), as it is difficult to estimate the expected above ground cosmic-ray exposure time for the C-4 crystals. An equivalent background Monte Carlo prediction for CoGeNT is also shown for reference (solid black line) \cite{longpaper}. Table~\ref{tab:backgrounds} contains estimates of each simulated contribution to the background. Reference \cite{longpaper} addresses the ability to fit and ``subtract'' the L-shell peaks based on corresponding K-shell peak heights at higher energies.}
\end{center}
\end{figure}

A simulation using Geant 4.9.4 was developed to estimate backgrounds and to verify the effectiveness of the shield design. In addition to the differences from CoGeNT in shield configuration and detector active volume, C-4 will employ two additional data analysis event selection criterion: (1) removing events having coincidences between the four germanium detectors and (2) removing events in coincidence with the muon veto (see \cite{longpaper} for a discussion of present use of the muon veto within CoGeNT). Table~\ref{tab:coincidences} summarizes the background reduction expected from rejecting events having coincidences between multiple detectors. Rejecting events having coincidences between detectors is particularly effective for the removal of neutron background signals. Figure~\ref{fig:comparison} compares the expected background in the 0.5-3.0~keVee energy region for C-4 with the measured background in the CoGeNT detector at SUL. Table~\ref{tab:dimensions} gives approximate dimensions used in the simulation. Estimates of the contributions to the total background from specific sources are presented in Table~\ref{tab:backgrounds}.
 
\begin{table}[ht!]
\begin{center}
\caption{\label{tab:coincidences}
The fraction of Geant4 simulated events having coincident energy depositions in multiple detectors (see text). For muon-induced events in the lead shield (``Pb shield: ($\mu$,n)'') the value (1) includes events due to muon-induced electromagnetic cascades and (2) does not include the application of the planned cosmic ray veto (which otherwise eliminates this background category, see Table \ref{tab:backgrounds}).}
\begin{tabular}{l c }
 & \\ \hline
Background & Coincidence \% \\ \hline
 & \\
\underline{$\beta$- and $\gamma$-ray backgrounds} & \\
Ge detectors: $^{3}$H activation & 0.02 \\
Ceramic resistors: $^{238}$U, $^{232}$Th & 1.7 \\
Cu cryostat/shield: $^{238}$U, $^{232}$Th & 2.1 \\
Pb shield: $^{238}$U, $^{232}$Th, $^{210}$Po & 2.8-4.1 \\
 & \\
\underline{Neutron backgrounds} & \\
Pb shield: ($\mu$,n) \emph{(see caption)} & 57 \\
HDPE: $^{238}$U fission & 26 \\
HDPE: $^{238}$U, $^{232}$Th ($\alpha$,n) & 22 \\
Cavern walls: ($\alpha$,n) & 40 \\
Cavern walls: ($\mu$,n) & \emph{n/a} \\
 & \\ \hline
\end{tabular}
\end{center}
\end{table}

\subsection{Simulation geometry}
The geometry used in the C-4 simulation is an approximation intended to study the propagation of radiation backgrounds through the shielding material. Some aspects of the shield design have been either simplified or removed. For example, the cryostats are approximated as simple cylinders. The OFHC copper dipsticks and four LN dewars have been omitted, as has a steel frame supporting the top polyethylene and water tanks. More detail has gone into the front-end electronics near the detectors, as it is found that the naturally occurring radioactivity of the resistors used there is a significant source of low-energy backgrounds \cite{longpaper}.

\begin{table*}[ht!]
\begin{center}
\caption{\label{tab:dimensions}
Dimensions of the C-4 shield simulation, taking $z$ as the vertical axis. The depth column refers to the global $z$ location of the center of the material. Thickness is then the amount of the material between the detectors and the external cavern. Support structures were omitted to simplify the geometry and decrease simulation time.}
\begin{tabular}{l c c c }
 & & & \\ \hline
Shield Layers (nested)   & Dimensions      & Depth & Thickness \\
(rectangular cuboids)    & (x, y, z) (cm)  & (cm)  & (cm)      \\ \hline
Water tank (4 sides)     & (327, 327, 247) & -124  & 51        \\
HDPE (Above shield)      & (226, 226, 61)  & -30.5 & 61        \\
HDPE (Below shield)      & (327, 327, 61)  & -278  & 61        \\
Muon veto (5 sides)      & (113, 113, 105) & -211  & 5         \\
Borated HDPE             & (103, 103, 100) & -158  & 2.5       \\
Aluminum                 & (97, 97, 97)    & -158  & 0.32      \\
Stock lead               & (91, 91, 86)    & -161  & 25.4      \\
Low background leads     & (40, 40, 38)    & -161  & 5.1       \\
OFHC copper cage         & (30, 30, 25)    & -161  & 0.32      \\ \hline
 & & & \\ \hline
Cryostat Interiors (4)   & Dimensions      & Depth & Thickness \\
(right cylinders)        & (r, z) (cm)     & (cm)  & (cm)      \\ \hline
OFHC copper cryostats    & (7, 22.9)       & -161  & 0.32      \\ 
Large ceramic resistors  & (0.3, 1.9)      & -165  & -         \\
Small ceramic resistors  & (0.2, 1.4)      & -165  & -         \\
Germanium detectors      & (4.5, 3.5)      & -172  & -         \\ \hline
 & & & \\
\end{tabular}
\end{center}
\end{table*}

\subsection{Muon-induced neutrons}
A significant reduction in background is expected from use of thick muon veto panels that will efficiently tag muon-induced neutron ($\mu$,n) scattering events for offline rejection. The C-4 shield will have 50.8~mm thick veto panels on five sides, with no veto panel on the bottom of the assembly. The veto panels are expected to offer a $\sim$99.7\% muon tagging efficiency at a triggering threshold of 7~MeV, sufficient to accept a majority of muon crossings while avoiding spurious environmental gamma triggers.  

%\begin{figure}[h!]
%\includegraphics[width=\columnwidth]{MuonVetoSpectrum.eps}
%\caption{\label{fig:vetoedep} Monte-carlo simulation of muon energy deposition in the C4 plastic scintillator veto panels for the average energy muons at the Soudan depth \cite{MeiHime}.  This is the summed energy from all 5 veto panels and the hump at around 20~MeV is from muons going through more than one panel.}
%\end{figure} 

To estimate neutron production inside the shield, muons are simulated as having the underground cosmic muon energy spectrum, angular distribution, and flux of $2 \times 10^{-7}$~cm$^{-2}$s$^{-1}$, taken from \cite{MeiHime} for the Soudan site.  The majority of energy depositions in the germanium detectors are from muon-induced neutrons, while $\sim$8\% of the veto-coincident events are from gammas and electrons produced in electromagnetic cascades as the muons lose energy in the shielding. Neutrons produced by muons passing through the cavern walls were simulated and normalized to a flux of $1.69 \times 10^{-8}$~cm$^{-2}$s$^{-1}$ \cite{MeiHime}. Neutrons often deposit energy in multiple detector volumes. Removing these multiple detector events from the simulated data set reduces the ($\mu$,n) rate originating in the shielding by 57\% and from cavern sources by 40\% (See Table \ref{tab:coincidences}). 

The present knowledge of muon fluxes underground and muon-induced production of neutrons generates significant uncertainties in these predictions, of the order of $\sim$20\% and $\sim$200\%, respectively. However, for C-4 these uncertainties are relatively unimportant, since the muon-veto will remove the vast majority of muon-related events, rendering any residual events from this source negligible compared to other sources of background.

\subsection{Fission and ($\alpha$,n) neutrons}
The flux of ($\alpha,n$) produced neutrons from radioactivity in the cavern rock, estimated at $3.8 \times 10^{-6}$~cm$^{-2}$s$^{-1}$ \cite{MeiHime}, is higher than the flux from muon spallation in rock, but with lower average energy. Gran Sasso ($\alpha,n$) neutrons are known to have a harder spectrum than Soudan~\cite{MeiHime}, and C-4 simulations use this as a worst-case estimate. The simulation shows neutrons from the cavern with energy lower than $\sim$5~MeV will not be able to traverse the full shield geometry in any significant number.

Based on assays performed at SNOLAB, the HDPE in the top and bottom outer layers of the C-4 neutron shielding is expected to have relatively high levels of $^{238}$U and $^{232}$Th contamination, 115$\pm$5~mBq/kg and 80$\pm$4~mBq/kg, respectively \cite{SNOLAB}. Though the C-4 design will utilize approximately 9.1~metric tons of stacked HDPE, simulations show that non-neutron radioactivity from the HDPE cannot penetrate the inner lead shield. However, $^{13}$C, which has a 1.07\% abundance in natural carbon, has a non-negligible cross section for the ($\alpha,n$) reaction at $\alpha$ energies found in the U and Th decay chains. The HDPE is therefore a source of ($\alpha,n$) neutrons. The neutron production from ($\alpha,n$) in HDPE has been calculated based on the cross-sections found in ENDF \cite{ENDF} and benchmarked with a SOURCES \cite{SOURCES-4A} calculation of ($\alpha,n$) on Teflon \cite{Perry}. Uranium-238 also has a small spontaneous fission branching ratio, with an average multiplicity per fission of 2.07 \cite{Axton}. The neutron events depositing energy in the C-4 region of interest from this spontaneous fission source are predicted to be a small fraction of the total neutron background rate.

\subsection{Cosmogenic $^{3}\!$H backgrounds in germanium}
Tritium decay is a concern for a low-energy Ge experiment since it is a $\beta$-decay with an end-point energy of only 18.6~keV. After germanium is zone refined and pulled into a crystal, the $^{3}$H concentration accumulates while above ground due to spallation from the fast neutron flux created by cosmic rays. Furthermore, $^{3}$H has a half-life of 12.33~years, allowing for only a small reduction over the lifetime of the experiment.  

This $^{3}$H decay was simulated within the detectors and normalized to the $^{3}$H production rate on surface given by \cite{Elliott} and \cite{Morales}. A two year surface exposure was assumed as a worst case scenario. Under this assumption, $^{3}$H is the second-largest contributor to the backgrounds in the 0-12~keV region of interest. This background can be directly reduced by limiting the time on surface of the germanium detectors, clearly a priority for the C-4 detector deployment.

\subsection{Radioactivity in resistors}
The ceramic used in many resistors has a relatively high concentration of $^{238}$U and $^{232}$Th. Such resistors reside in the front-end electronic read-out of commercial $p$-type point contact detectors. Simulations have shown it is likely the flat background in the 0-12~keV region of the CoGeNT energy spectrum is due to these front-end resistors \cite{longpaper}. There are several possible methods of redesigning the cryostat assembly to reduce this background in C-4. The simplest approach is removing a large resistor used for heating the front-end field-effect transitor (FET), relying on FET self-heating instead. Based on past experience with high-quality FETs, this should be possible without paying an excessive penalty in electronic series noise (Section \ref{sec:noise}). In each cryostat, this leaves only one resistor with an approximate mass of 20 mg, located approximately 4.8~cm from the germanium detector. We are presently studying an approach to its additional removal without loss of performance. Simulations indicate that the removal of the heating resistor decreases the contributions from $^{234}$Pa, $^{228}$Ac, $^{208}$Tl, $^{212}$Pb, $^{214}$Bi, and $^{214}$Pb gammas in the $^{238}$U and $^{232}$Th chains by 75\%. Despite this reduction, naturally occurring radioactivity in the electronics still contributes significantly, with the disadvantage that coincidence tagging between detectors will only remove $<$5\% of these events. Developments in low-background front-ends such as those for the \textsc{Majorana Demonstrator} \cite{MJD,LMFE} could further reduce this background in C-4.

\subsection{Radioactivity in the lead shield and OFHC copper}
Contributions to the C-4 background from radioactivity in the outer lead shield come from gamma-rays produced in the U and Th decay chains. The largest contributors are $^{234}$Pa, $^{228}$Ac, $^{208}$Tl, $^{212}$Pb, $^{214}$Bi, and $^{214}$Pb. The outer lead shield was simulated as having concentrations of 0.41~mBq/kg and 0.08~mBq/kg of $^{238}$U and $^{232}$Th, respectively \cite{SNOLAB}, and the decay chains were assumed to be in secular equilibrium. A concentration of 0.02~Bq/kg $^{210}$Pb was used for the low-background inner lead, measured using radiochemical extraction followed by alpha spectroscopy at PNNL \cite{smiley}. Despite a high assumed $^{210}$Pb concentration of $\sim$93~Bq/kg in the outermost lead layer \cite{SNOLAB}, the inner low-background lead effectively prevents $\beta$-decay bremsstrahlung photons from reaching the detectors \cite{longpaper}.

In these background simulations, high purity OFHC copper is used for the cryostat and a copper cage inside the lead shield in the C-4 design. All OFHC copper was simulated as having concentrations of 18~$\mu$Bq/kg and 28~$\mu$Bq/kg for $^{232}$Th and $^{238}$U \cite{Hoppe}, respectively, and the decay chains were assumed to be in secular equilibrium. Both the beta and gamma emitters in the $^{232}$Th and $^{238}$U decay chains were included in the simulation. Results are presented in Table~\ref{tab:backgrounds}. Note that use of ultra-high purity electroformed copper in place of OFHC copper during actual fabrication would reduce these backgrounds and is ideal for forming the cylindrical cryostat end caps.

\begin{table*}[ht!]
\begin{center}
\caption{\label{tab:backgrounds}
Summary of simulated backgrounds in the C-4 experiment compared to CoGeNT. For muon-induced events in the lead shield (``Pb shield: ($\mu$,n)''), (1) the rates include events due to muon-induced electromagnetic cascades and (2) the CoGeNT rate does not have a cosmic ray veto applied while the C-4 rates assume the planned cosmic ray veto is active.}
\begin{tabular}{l c c c}
 & & & \\ \hline
Backgrounds (counts/kg/day) & CoGeNT & C-4 & C-4 \\
  & (0.5-3 keV) & (0.5-3 keV) & (0.5-12 keV) \\ \hline
 & & & \\
\underline{$\beta$- and $\gamma$-ray backgrounds} & & & \\
Ge detectors: $^{3}$H activation & 8.3$\times 10^{-1}$ & $\phantom{<}$8.2$\times 10^{-1}$ & $\phantom{<}$2.8$\times 10^{0\phantom{-}}$ \\
Ceramic resistors: $^{238}$U, $^{232}$Th & 6.2$\times 10^{0\phantom{-}}$ & $\phantom{<}$1.4$\times 10^{0\phantom{-}}$ & $\phantom{<}$6.5$\times 10^{0\phantom{-}}$ \\
Cu cryostat/shield: $^{238}$U, $^{232}$Th & 5.0$\times 10^{-2}$ & $\phantom{<}$7.1$\times 10^{-3}$ & $\phantom{<}$3.4$\times 10^{-2}$ \\
Pb shield: $^{238}$U, $^{232}$Th, $^{210}$Po & 3.4$\times 10^{-1}$ & $\phantom{<}$1.4$\times 10^{-1}$ & $\phantom{<}$5.5$\times 10^{-1}$ \\
 & & & \\
\underline{Neutron backgrounds} & & & \\
Pb shield: ($\mu$,n) \emph{(see caption)} & 1.9$\times 10^{0\phantom{-}}$ & $\phantom{<}$2.6$\times 10^{-4}$ & $\phantom{<}$6.4$\times 10^{-4}$ \\
HDPE: $^{238}$U fission & 1.2$\times 10^{-4}$ & $\phantom{<}$5.7$\times 10^{-5}$ & $\phantom{<}$1.3$\times 10^{-4}$ \\
HDPE: $^{238}$U, $^{232}$Th ($\alpha$,n) & 2.2$\times 10^{-4}$ & $\phantom{<}$1.3$\times 10^{-4}$ & $\phantom{<}$3.3$\times 10^{-3}$ \\
Cavern walls: ($\alpha$,n) & 3.0$\times 10^{-1}$ & $<$1.4$\times 10^{-3}$ & $<$1.4$\times 10^{-3}$  \\
Cavern walls: ($\mu$,n) & 7.7$\times 10^{-3}$ & $\phantom{<}$6.7$\times 10^{-3}$ & $\phantom{<}$1.6$\times 10^{-2}$ \\
 & & & \\ \hline
Estimated total rate & 10 & 2 & 10 \\ \hline
\end{tabular}
\end{center}
\end{table*}

\section{Improvements to PPC electronic noise and energy threshold}
\label{sec:noise}

The CoGeNT collaboration has operated several PPCs in ultra-low background environments, with special attention paid to reduction of environmental EM/RF noise and monitoring of long-term detector stability. This has allowed us to discern several limitations still affecting the performance of this germanium detector design. For instance, we have experimented with the choice and packaging of the first-stage of amplification (e.g., the FET) in an attempt to identify the origin of residual sources of electronic noise still limiting the ultimate energy resolution and the lowest energy detectable with these devices. 
 
\begin{figure}[ht!]
\begin{center}
\includegraphics[width=\columnwidth]{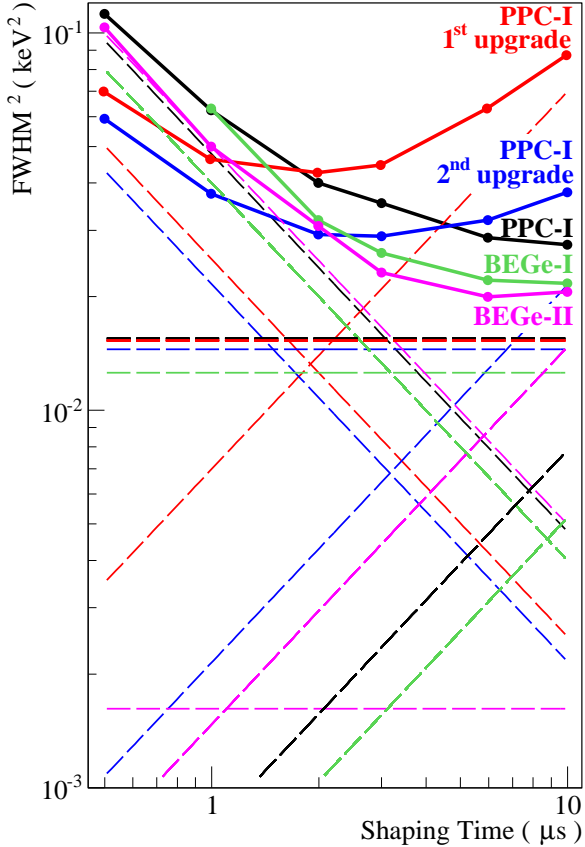}
\caption{\label{fig:Noise} Electronic noise contributions measured with a pulser, for a number of PPC detectors and their upgrades. The (flat) non-white component remained mostly independent of the sum of detector and FET capacitance until BEGe-II, pointing at a dominant parallel-f noise component (see text).}
\end{center}
\end{figure}

Figure \ref{fig:Noise} encapsulates the results from these studies. As is customary \cite{BertuccioRevSci}, detector noise is plotted versus shaping time in the second stage of amplification (shaping amplifier), decomposing it into its three main components: series, parallel, and non-white (the sum of ``parallel-f'' and ``series-1/f'' noise). Measurements are labelled according to PPC detector tested. ``BEGe'' makes reference to ``Broad Energy Germanium'', a commercial quasi-planar PPC geometry available from Canberra Industries. We briefly list here our observations of the noise in CoGeNT PPC detectors organized by each type of electronic noise.

\textit{Series noise} (inversely proportional to shaping time): This component is the most predictable of the three and is presently dominated by the total capacitance ($C_{in}$, sum of detector plus FET capacitances) of the device. The variations visible in Figure \ref{fig:Noise} change according to the capacitance of the tested FET ($C_{FET}$), for a PPC contribution in the narrow range $C_{DET} = $~2-3~pF, secondarily depending on FET operating temperature. We have explored the range $C_{FET} = $~0.9-4.8~pF using several FETs manufactured by Moxtek (MX20, MX120, MX120E) and Canberra (SF202, EuriFETs 106 and 102d), some of these FETs being under development, observing the predictable variation \cite{BertuccioNIMA} for this component of the noise. A further reduction of this non-dominant component can be obtained via optimization of the preamplifier, choice of pulsed-reset mechanism, and a further reduction of detector capacitance through planned modifications to the contact geometry.
 
\textit{``Flat'' or non-white noise} (independent of shaping time): Several know sources for this component \cite{BertuccioNIMA} have been investigated. Chief among these is the packaging of the FET; in commercial versions it can involve sub-optimal dielectrics or other sources of capacitance liable to generate 1/f contributions of this type. As is visible in Figure \ref{fig:Noise}, we observed a dominant non-white component that was not removed after several improvements to the standard FET package, for instance, replacement of boron nitride FET capsules by polytetrafluoroethylene (PTFE) equivalents.
 
A tell-tale feature is the approximate independence of this component on the value of $C_{in} = C_{FET} + C_{DET}$. This is characteristic of the so-called ``parallel-f'' electronic noise \cite{radeka}. Inspection of the mounting of the germanium crystals within their cryostats revealed the use of dielectrics forming lossy paths to ground in parallel to the detector element, with sufficient parasitic capacitance to explain the bulk of this limiting flat component. Their contribution to the noise budget was calculated as in \cite{BertuccioNIMA}, using as input the parasitic capacitance values measured directly on inner cryostat replicas. These calculations provided a good quantitative agreement with the observed flat noise component, confirming a parallel-f origin. The final detector prepared (BEGe-II) modified the PTFE liner used to insulate the crystal in the cryostat in an attempt to impact this parallel-f contribution. The drop seen has informed the cryostat design for the C-4 cryostats.

The internals of C-4 detectors have been redesigned relying on ANSYS Maxwell (``MAXWELL-3D'') \cite{maxwell3d} simulations of parasitic capacitance affecting the gate of the FET \cite{c4cryostat}. A considerable abatement of this source of electronic noise is expected from this custom C-4 crystal mount design. This should result in an improvement to the presently achievable $\sim$0.5~keVee PPC energy threshold, leading to additional sensitivity to low-mass dark matter candidates. However, we remark that noise components add in quadrature and that sub-dominant components may lurk not far below this parallel-f component. This leads us to be cautious in our expectations. Furthermore, these cryostat design modifications are not trivial to implement as they can affect and inhibit the thermal cooling of the crystal, resulting in higher temperatures at the detector element that may result in increased leakage current and parallel noise. Modifications for increasing the thermal cooling power using a custom cryostat cold finger are readily achievable with little or no impact on the background estimates.

\textit{Parallel noise} (directly proportional to the shaping time): The large variation of values observed for this component in Figure \ref{fig:Noise}  is dominated by the magnitude of the leakage current in the detectors \cite{BertuccioNIMA}, the lowest value obtained (BEGe-I, the CoGeNT detector presently running at SUL) corresponding to $\sim$0.9~pA. The exact preparation of the intercontact surface (wet chemistry, passivation) is critical in order to obtain the lowest possible value of this component and to ensure its stability over time. The techniques presently used in detector manufacture, while sufficient for conventional higher-capacitance (higher-noise) germanium detectors, have not consistently yielded the desired small leakage currents at the standard detector temperature of $\sim$90~K. Much trial-and-error is still involved in the optimization of leakage current during PPC detector fabrication, in order to achieve levels low enough not to dominate the noise budget. Our industrial partner, Canberra Industries, is presently addressing this issue via systematic studies of intracontact surface passivation, aiming at consistently obtaining stable $<$1~pA leakage currents in all PPCs. 

Within the C-4 program, we will attempt a comprehensive effort to simultaneously reduce all of the remaining sources of electronic noise in PPCs. While the presently achieved PPC energy threshold is already an order of magnitude lower than typical values for standard coaxial germanium diodes, our ambitious goal is to ultimately match the state-of-the-art for small silicon X-ray detectors with $C_{DET} \sim $1~pF resulting in an energy threshold of $\sim$50~eV. It is important to remark in this respect that detector grade germanium contains no appreciable concentration of trapping and recombination centers. We therefore neither expect nor have observed any noise-generating processes arising from bulk material. For example, BEGe-II (Figure \ref{fig:Noise}) has a mass of $\sim$800~g -- almost twice that of the three PPCs previously tested -- with no deleterious observable effect on the noise related to crystal mass.

\section{Scientific Reach}
\label{sec:science}

We explore two scenarios when studying the dark matter sensitivity of the C-4 dark matter experiment. The first case is C-4's exclusion reach in WIMP mass and cross-section parameter space, if the experiment were to fail to observe the low-energy excess observed by CoGeNT \cite{longpaper}. This sensitivity is calculated based on the background budget predicted in Section \ref{sec:backgrounds}. The second case is the sensitivity of C-4 to a WIMP signal having mass and cross-section parameters compatible with a dark matter interpretation of previous CoGeNT results. The following two sub-sections explore the C-4 experimental sensitivity under these two premises.

\subsection{WIMP Sensitivity}
\begin{figure}[ht!]
\begin{center}
\includegraphics[width=\columnwidth]{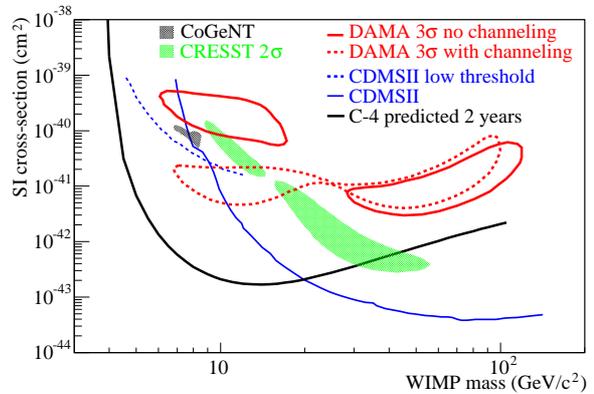}
\caption{\label{fig:c4exclusion} The predicted C-4 WIMP sensitivity after 2~years of running compared to CDMS II sensitivity curves. C-4 is predicted to offer a considerably better sensitivity than the low-threshold CDMS II analysis, up to a WIMP mass of $\sim$15~GeV/c$^{2}$. We conservatively do not assume any of the expected improvements to the energy threshold discussed in \ref{sec:noise} when calculating these projections.}
\end{center}
\end{figure}

As discussed in Section \ref{sec:backgrounds}, the expected C-4 background is flat over the analysis region 0.5-3.0~keVee. Figure ~\ref{fig:c4exclusion} shows the 90\% limit to spin-independent WIMP couplings as a function of WIMP mass assuming this predicted $\sim$5$\times$ reduction in background compared to CoGeNT. If the observed C-4 spectrum is flat as in these predictions, C-4 will offer a very competitive sensitivity to low-mass WIMPs, compared to other current experiments. Figure \ref{fig:c4exclusion} also shows the CDMS II low-threshold and combined Soudan WIMP sensitivity \cite{CDMSLOW, CDMSOUDAN}. C-4 operation with a 4.1~kg target mass and 8~months of data collection would exclude a cross-section down to 10$^{-42}$~cm$^{2}$ for 10~GeV/c$^{2}$ WIMPs, more than an order of magnitude below the region of interest generated by the CoGeNT result. We conservatively do not assume any of the expected improvements to the energy threshold discussed in Section \ref{sec:noise} when calculating these projections.

%\begin{figure}[htp]
%\begin{center}
%\includegraphics[width=\columnwidth]{limitPlotTime.eps}
%\caption{\label{fig:c4runtime} Predicted sensitivity of C4 to 7 and 10~GeV/c$^{2}$ WIMPs as a function of months of running at 4.1~kg total mass.}
%\end{center}
%\end{figure}

\subsection{Annual Modulation sensitivity}
\begin{figure}[ht!]
\begin{center}
\includegraphics[width=\columnwidth]{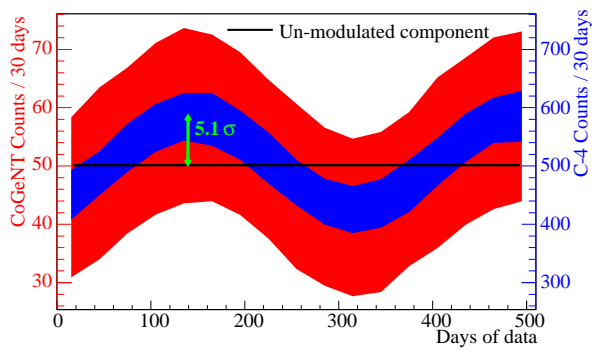}
\caption{\label{fig:c4modpred} The predicted sensitivity of C-4 (inner narrow blue band - right-hand scale) to an annual modulation assuming the observed CoGeNT modulation amplitude (wider red band - left-hand scale) is due to WIMP dark matter interactions (see text).}
\end{center}
\end{figure}

An analysis of 15 months of CoGeNT data \cite{cogentPRL107} indicates the possible presence of a modulation in low-energy bulk event rate, compatible in principle with an annual modulation signature expected from dark matter interactions \cite{Drukier86}. The best-fit modulation amplitude determined in \cite{cogentPRL107} was $\sim$16.6$\pm$3.8\%, with a rejection of the no-modulation hypothesis at 2.8$\sigma$ confidence level. The increased mass of the C-4 target will enable a confirmation or rejection of the observed CoGeNT modulation. Figure \ref{fig:c4modpred} offers an estimate of the improved sensitivity of C-4 to an annual modulation. The wider red band in Figure~\ref{fig:c4modpred} represents the CoGeNT modulation \cite{cogentPRL107} with 3$\sigma$ uncertainties, and the inner narrow blue band is the predicted C-4 modulation, also at 3$\sigma$. The narrowing of the C-4 inner narrow blue-band is mainly due to the 10$\times$ increase in active target mass, leading to a reduction in statistical uncertainties. If the CoGeNT modulation has a dark matter origin, C-4 should be able to observe it at $\sim$5$\sigma$ statistical significance within a year of operation.

\section{Conclusions}
\label{Conclusions}

\begin{figure}[ht!]
\begin{center}
\includegraphics[width=\columnwidth]{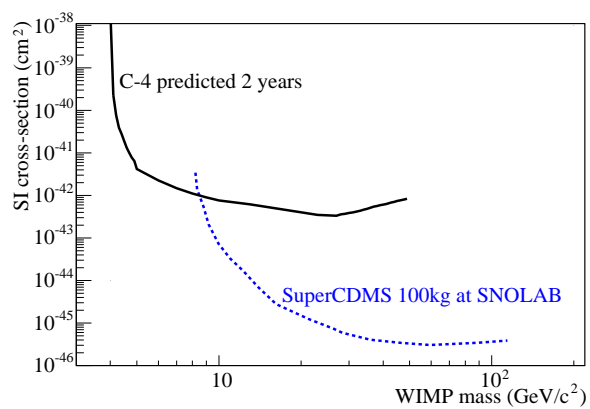}
\caption{\label{fig:scdms} Comparison of the C-4 and SuperCDMS projected exclusion regions for low-mass WIMP candidates (see text).}
\end{center}
\end{figure}

The relatively simple technology employed by C-4, p-type point contact germanium detectors, allows for a very low energy threshold, excellent energy resolution, and good surface event rejection. This style of detector is however unable to distinguish nuclear recoil signals from electron recoil backgrounds, having to rely instead on background abatement though material selection and shielding. The CoGeNT detector has demonstrated a remarkable long-term stability, essential in searches for an annual modulation dark matter signature \cite{longpaper}. C-4 will increase the germanium target mass by a factor of ten, while reducing backgrounds and matching or improving on CoGeNT's energy threshold. In the absence of a dark matter signal, the ability to operate with a live time fraction of nearly 100\% \cite{longpaper}, will allow C-4 to quickly (2 years) accumulate an exposure sufficient to exclude low-mass WIMP candidates ($\sim$10~GeV/c$^{2}$) down to a coupling of $\sim7\times10^{-43}$~cm$^{2}$. This compares well with projections for SuperCDMS (100~kg) operation at SNOLAB \cite{SuperCDMS}, as shown in Figure~\ref{fig:scdms}. The increased sensitivity of C-4 to low-mass dark matter will confirm or definitively exclude a dark matter origin for the anomalies registered by the DAMA, CoGeNT, and CRESST experiments. If the CoGeNT modulation result persists and is due to dark matter interactions, C-4 should identify this modulation with 5$\sigma$ significance within a year of operation.

\section{Acknowledgments}
\label{Acknowledgments}

The Ultra-Sensitive Nuclear Measurement (USNM) Initiative, a Laboratory Directed Research and Development (LDRD) program at the Pacific Northwest National Laboratory partially supported this work. The authors thank the NSF (grants PHY-0653605 and PHY-1003940) and the Kavli Foundation for partially supporting this work. N.E. Fields is supported by grant DE-FC52-08NA28752 from the DOE/NNSA Stewardship Science Graduate Fellowship program. T.W. Hossbach is partially supported by the Intelligence Community (IC) Postdoctoral Research Fellowship Program.

%\section*{References}



%% References

%\bibliographystyle{elsarticle-num}
%\bibliography{bib}





% \end{thebibliography}

\end{document}